# Hiding Data in Images Using Cryptography and Deep Neural Network

**Kartik Sharma[1, a, ]***, **Ashutosh Aggarwal[1, b]**, **Tanay Singhania[2, c]**, **Deepak Gupta[1, d]**,
**Ashish Khanna[1, e]**

[1] Computer Science and Engineering Department, Guru Gobind Singh Indraprastha University, New Delhi, India
[2] Applied Mathematics Department, Delhi Technological University, New Delhi, India
[a.] 98kartik.sharma@gmail.com, [b.] ashutoshaggarwal98@gmail.com, [c.] tanay_bt2k17@dtu.ac.in, [d.] deepakgupta@mait.ac.in,
[e.] ashishkhanna@mait.ac.in
*Corresponding Author: Kartik Sharma

## Abstract

Steganography is an art of obscuring data inside another quotidian file of similar or varying types. Hiding data has always been of significant importance to digital forensics. Previously, steganography has been combined with cryptography and neural networks separately. Whereas, this research combines steganography, cryptography with the neural networks all together to hide an image inside another container image of the larger or same size. Although the cryptographic technique used is quite simple, but is effective when convoluted with deep neural nets. Other steganography techniques involve hiding data efficiently, but in a uniform pattern which makes it less secure. This method targets both the challenges and make data hiding secure and non-uniform.

## 1. Introduction

The ingenuity of enshrouding a file within a different file is called Steganography that has been in exercise from the past 2500 years. Better steganography definitions are given by Provos and Honeyman [1]; Johnson and Jajodia [2]; Morkel et al. [3]; Fridrich and Goljan [4]. Predominantly, in steganography, there is a secret message and a carrier and the arcane message is hidden in the carrier in some way that it is undetectable. Therefore, it has multiple applications in defense organizations, intelligence agencies, identity cards, etc.

Steganography is mainly of 4 types; image, text, audio/video and protocol which are well explained by Shih [5] and Morkel et al. [3].

The method used to recover the message hidden by steganography is called Steganalysis.

Steganalysis is lucidly explained by Shih [5]; Fridrich and Goljan [4]; Avcibas et al. [6]. Hitherto, steganography was effortless to implement securely as people did not know much about it. But with the headway in the Steganalysis, encrypting the secret message before passing it to a carrier has become a prerequisite. This paper mainly focuses on Image steganography as it is one of the most vastly used techniques.

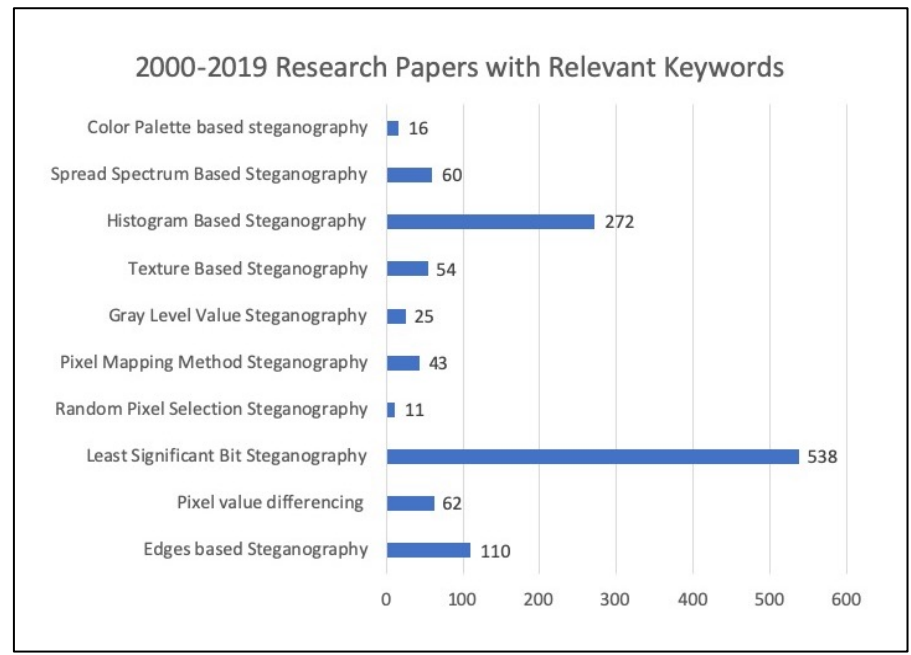

**Figure 1.** Research Papers (2000-2019) with relevant Keywords.

Now, there are profuse techniques for Image Steganography as explained by Li et al. [7]; Subhedar et al. [8]. The bar graph plotted in Figure 1 exhibits the frequency of different techniques used. Out of these techniques, the most common method is the Least Significant Bit Steganography. Therefore, this habitual and widely accepted method is used. Moreover, the Least Significant Bit Cryptography easily dovetails with Deep Neural Networks, which is explained further in this paper.

In LSB Steganography, data is hidden in the least significant bit of the carrier [9], [10]. While talking about images in specific, alteration is performed on the last bit from the 8-bit carrier pixel value. Due to changing only the least significant bits of the cover image, it goes unnoticed to the naked eye that there is a slight change in the color values. A few years back, it was a good practice, but as the method disseminated, Steganalysts found a method to recover it, which made it no longer usable. Secret image can be easily found by extracting the least significant bits of the cover image.

Different tools such as ZSteg, JSteg, StegoVeritas, etcetera. can be used to find the Least Significant bit values. As using the LSB technique became trite, Baluja [11] implemented its variant with Deep Neural Networks to make it secure. The method proposed by Baluja [11], had a constraint that the image could not be made public. Addressing this specific constraint, an improved algorithm involving cryptography and deep neural networks is discussed in the paper.

Theoretical, as well as the practical implementation of the algorithm, is discussed in the paper. The algorithm is trained on a public dataset, producing an accuracy of 90% approximately on both training and test set, therefore making it an effective algorithm for data hiding.

The paper proposes the following:

- An improved algorithm for Steganography using Cryptography and Deep Neural Networks is propounded for hiding data.
- This algorithm uses Convolutional Neural Networks with Adam optimizer for both hiding and revealing part of the architecture.
- Public Dataset is used for training this algorithm, making the accuracy of approximately 90% on both train and test dataset.

The remaining paper is structured as follows. The second section includes reviews of varying papers. The third and fourth section contains the methodology and results respectively. Section 5 explains the advantages of using this algorithm. Section 6 and 7 describe conclusion and future work respectively.

## 2. Literature Review

The paper propounded by Morkel et al. [3] focused mainly on Image steganography while briefly explaining other steganography techniques. The paper proposed various techniques to hide data in images and a detailed comparison of various steganography techniques.

Li et al. [7] provided a survey on Steganalysis and steganography for images. They mainly used JPEG format images for their research and provided extensive research explaining different steganographic techniques as well as the attack scenarios on those methods. Different tools such as OutGuess, JSteg/JHide, and methodologies such as LSB, LSB Matching, Stochastic steganography, etc. were used for demonstration.

Anderson et al. [12] explained various limitations in image-based steganography when using these theoretical concepts privately. Therefore, they proposed public-key steganography with parity checks to amplify covertness. Public key steganography was demonstrated using passive and sometimes with the active warden as well.

Marwaha and Marwaha [13] used Asymmetric Key Cryptography to encrypt the message. Specifically, the Data Encryption Standard (DES) was used to encrypt the data. Furthermore, the cipher was hidden in the image file. This technique worked only when the container image was transmitted in a lossless environment. Data Encryption Algorithm (DEA) was used to extract the secret text back from the image.

Baluja [11] implemented the Least Significant Bit steganography using convolutional neural networks. He was successfully able to hide and reveal the image. However, his research has a downside that if the cover image is made public, the secret image could be extracted from the cover by amplifying the difference.

Research presented by Djebbar et al. [14] showcased various audio steganography techniques. Different methods such as low bit encoding, magnitude spectrum, wavelet coefficients, bitstream hiding, etc. were evaluated in several hiding domains such as temporal, frequency, etc. They compared the methods mentioned earlier on the bases of hiding rate, capacity available, distortions, robustness, and lossy compression.

Biometric data protection was well explained by Douglas et al. [15]. They explained various biometric techniques such as fingerprint, retina, face and their issues. These different biometric were further used to hide data using some well-known image steganography methods like Spatial Domain Techniques (LSB-Steganography), Pixel Value Differencing (PVD), Transform Domain Techniques (DCT/ DWT) and some hybrid techniques such as Singular Value Decomposition (SVD). Binarization was incorporated to hide data inside biometric data.

Jiang et al. [16] proposed a new LSB based Quantum Steganography Algorithm for images. The two algorithms presented were advantageous as they as they were absolutely blind as well as no classical computers were required to perform this. Image preprocessing was done using the Hilbert image scrambling algorithm that permuted the pixels into new positions.

Wang et al. [17] briefly explained various steganography techniques and tools. Different steganalysis techniques to break the above mentioned steganographic methodologies such as POV-based Chi-Square test, Palette Checking, Universal Blind Detection, etc. were succinctly expounded.

## 3. Methodology

### 3.1. Artificial Neural Network

Artificial Neural Network (ANN) is a progression of an algorithm that behaves like a human sensory system [18]. An ANN embodies a huge number of interconnected processing units called nodes (neurons) that work collaboratively to solve a fastidious errand. In a similar fashion to our brain, the nodes work in parallel and each node collects input from others. The nodes then compute these inputs and pass on the data to the next connected node. Neural networks learn to perform tasks by analyzing pre-defined data.

Convolutional Neural Network (CNN) is one of the widely used algorithms for Neural Networks. CNN or ConvNet is mostly used to analyze images. CNN's allow us to encode image-specific features into the architecture, making the network more fitting for the tasks, handling the images - whilst reducing the parameters required to set up the model. A better understanding can be found in [19], [20].

### 3.1.1. Convolutional Neural Network

CNNs are very powerful neural networks that can be classified as a regularized version of Multilayer Perceptrons or MLP [21]. The name "convolutional neural network" indicates that the network employs a mathematical operation called convolution [22]. Convolution is a specific kind of linear operation that expresses the amount of overlap of one function as it is shifted over another function. Convolutional networks are simply neural networks that use convolution in place of general matrix multiplication in at least one of their layers [23]-[25]. Reported studies [26]-[28] suggest that CNN based architecture for encoding and decoding features have the following advantages:

- CNN extract features of the images automatically.
- CNN effectively uses adjacent pixel information to effectively downsample the image first by convolution and then uses a prediction layer at the end.
- CNN performs well and gives better accuracy.

Using a deep neural network, CNN, in this case, one will be able to have a good sense of the patterns of natural images. The network will be able to make judgments on which area is redundant, and more pixels can be hidden there. By conserving the space on unnecessary areas, the amount of hidden data can be increased. Since the structure and the weights can be randomized, the network will hide the data that cannot be known to anybody who doesn't have the weights [29], [30].

## 3.2. Architecture

The network architecture is somewhat similar to Auto-Encoders [31]. In general, Auto-Encoders are used to reproduce the input ensuing a set of transformations. By performing this, they learn about the characteristics of the input distribution.

But in our illustration, the proposed architecture is marginally distinct. Rather than simply procreating images, the network has to conceal an image, additionally procreate another image. The proposed architecture is exhibited in Figure 2.

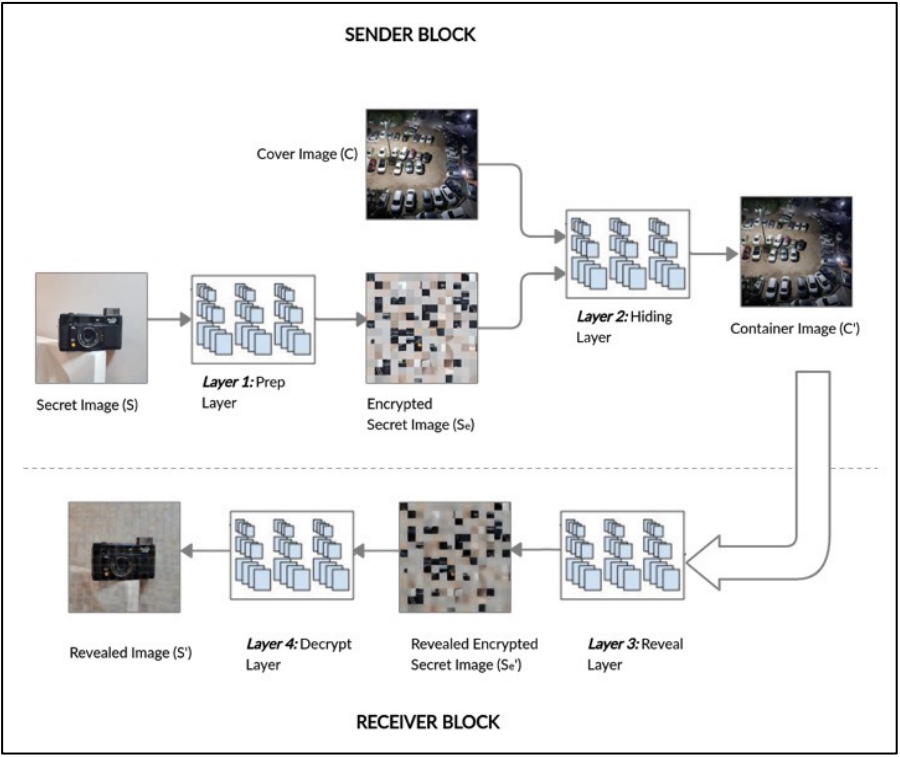

**Figure 2.** Architecture of the proposed system. Top: Sender Block. Bottom: Receiver Block

The four components shown in Figure 2 are trained as a single network but it will be easier to explain them individually. The first layer, Prep Layer, prepares the secret image to be hidden. This layer serves various purposes. Firstly, if the secret image is smaller than the cover image, it increases the size of the secret image to the size of the cover image, hence distributing the secret image's bits across the entire pixels. Furthermore, relevant to all measurements of hidden images, the purpose is to reconstruct the color-based pixels to more useful features for succinctly encoding the image such as edges. The most important purpose of this layer was to finally embed the secret image to the encrypted image so as to avoid any leak of the secret message.

The second layer, Hiding Layer, takes the output of the Prep Layer as input and a cover image so as to produce the Container Image. The input to this network is a square pixel image, with depth considered of RGB channels of the cover image and the reconstructed channels of the secret image. These two layers together form the Sender Block. The Container image produced can be shared with the receiver.

The third Layer, Reveal Layer, is used by the receiver to produce back the encrypted image. This layer takes in the Container image as input and removes the cover image to generate the encrypted secret image.

The fourth Layer, Decrypt Layer, takes in the output of the Reveal Layer and decrypt the image to finally reveal the secret image. The third and fourth layer together forms the Receiver Block.

As shown by Baluja [11], the loss in the network can be computed by the given formula:

$$L\left(c,c',s,s'\right)=\left\|c-c'\right\|+\beta\left\|s-s'\right\|$$ (1)

where the variables are as mentioned in Figure 2. The loss is the conventional Mean Square Error between the original cover image and the container image, and β*(MSE between the original secret image and the revealed image). β is a hyper-parameter that regulates how much the secret image should be restored. Since the given function is differentiable, the neural network can be trained end-to-end.

### 3.3. Optimizer

The role of the optimizer is to improve the weight parameters to minimize the loss function. It leads the loss function to find its global minima. There are various types of optimizers, for example - Momentum, Nesterov Accelerated Gradient, Adagrad, Adadelta, Adam, etc.

The model uses ADAM - Adaptive Moment Estimation [32] optimizer that calculates a separate learning rate for each parameter. It is computationally productive and has very low memory requirements which make it optimum for our model. The reason behind not using Momentum, Stochastic Gradient Descent, and Nestorov optimizers is because the dataset used is sparse. Other adaptive learning methods such as RMSprop, Adagrad, and Adadelta were outperformed by the ADAM optimizer making it the ideal choice.

ADAM computes the mean and uncentered variance of past gradients as $p_t$ and $q_t$ with β1, β2 ∈ [0, 1) as hyper-parameters respectively.

$$p_t = \beta_1 p_t - 1 + \left(1 - \beta_1\right) g_t \qquad (2)$$

$$q_t = \beta_2 p_t - 1 + \left(1 - \beta_2\right) g^2{}_t \qquad (3)$$

Further, it computes bias-corrected first moment ($p_t$) and second moment ($q_t$) as follows:

$$\hat{p}_t = \frac{p_t}{1 - \beta_1^t} \qquad (4)$$

$$\hat{q}_t = \frac{q_t}{1 - \beta_2^t} \qquad (5)$$

Then these bias-corrected mean and variance of past gradients are used to update the parameters of the model as follows:

$$\theta_{t+1} = \theta_t - \frac{n}{\sqrt{\hat{q}_t} + \varepsilon} \hat{p}_t \qquad (6)$$

where $\varepsilon$ is a small scalar used to prevent division by zero and $g_t$ refers to the loss function of the model.

## 3.4 Image Encryption

Images are widely used as part of communication since they are more effortless to process than text. As explained by Kumar et al. [33], the user might communicate on a network that is compromised and security of the information becomes crucial in such cases. Hence, encryption or the scheme used to transform an image into another image that is not easily comprehensible [34], [35] is important to maintain the security of the information being communicated. Image encryption has applications in various areas - for example, military, security agencies or wherever the data is sensitive or contains privileged information.

Image Encryption methods or algorithms revolve around the following three ideas:

- Pixel Permutation - Scrambling the pixels
- Pixel Substitution - Modifying each pixel value
- Visual Alteration

An image histogram is one of the security parameters kept in consideration for encryption methods as explained by Li et al. [36]. The histogram displays the frequency distribution and provides an insight into the frequency of each pixel value. In the various encryption methods as explained by Khan and Shah [37], methods based on the transformation of pixel values for example - AES (Advanced Encryption System) tend to create a uniform image histogram that protects it from plain text attack but requires zero loss while decrypting the image. This is not useful in this case as the neural network used (to hide images in other images) depends on the redundant area in the cover images which in turn gives the amount of hidden data. Similarly, ECC (Elliptic Curve Cryptography) as explained by Geetha et al. [38] cannot be used in our scenario.

Due to the data loss in the neural networks, methods like AES and ECC fail. Hence, one can use methods based on scrambling of data (in this case pixels) retaining the same pixel value but at a different location giving the same histogram [39]. The scrambling helps in providing the same histogram but decreases the correlation between pixels.

In this paper, scrambling or shuffling blocks of pixels technique is used [40]. As discussed by Kumar et al. [33], displacement techniques give the same image histogram but a reduced correlation between adjacent pixels of the image as illustrated in Figure 3.

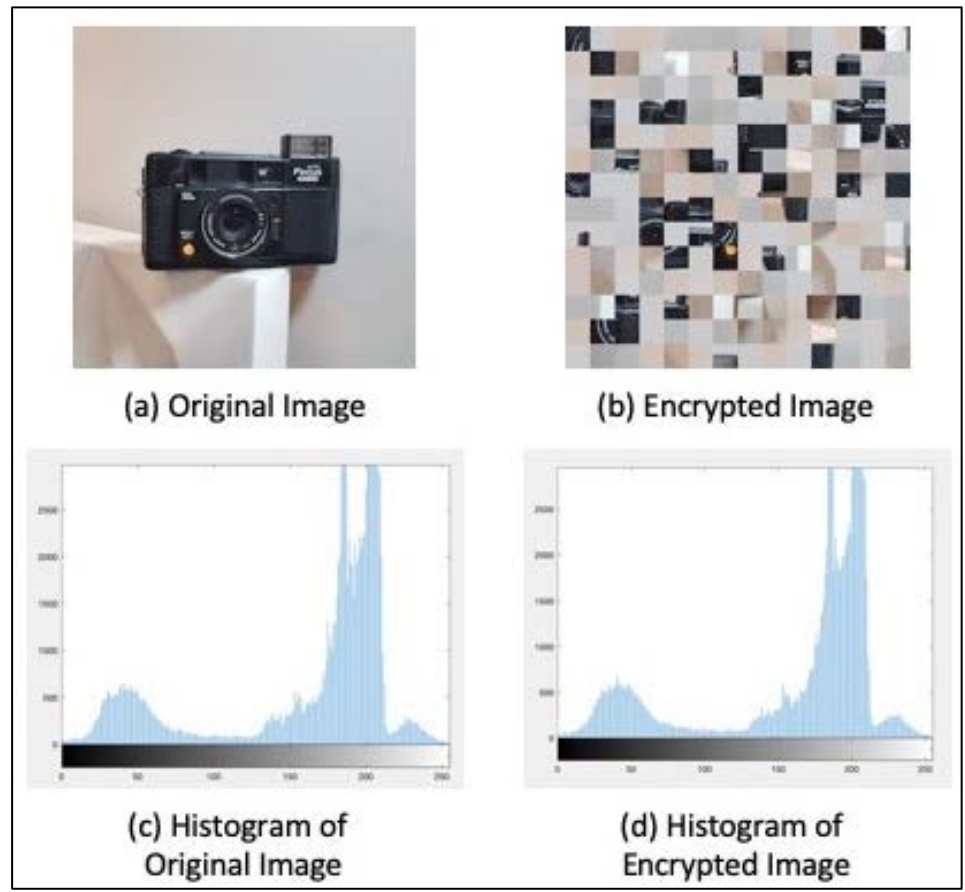

**Figure 3.** Histogram comparison of the original and encrypted image

As the number of blocks increases, the correlation decreases giving a suitable encryption layer. The number of blocks can be referred to as the order of encryption. By increasing the order of encryption it is observed that one cannot interpret the significance of the image just by looking at it. Figure 4 depicts the effect of the increasing order of encryption.

After experimenting, it is observed that the order of encryption 196 gives the most suitable level of encryption and low correlation. Moreover, 196 blocks of pixels mean 196! ways to arrange them. This gives $\cong 5.08 \times 10365$ number of permutations and considering a computer does 1016 calculations per second then it will take about $\cong 10342$ years to find the right permutation. Hence it is used as the first layer of encryption in the network.

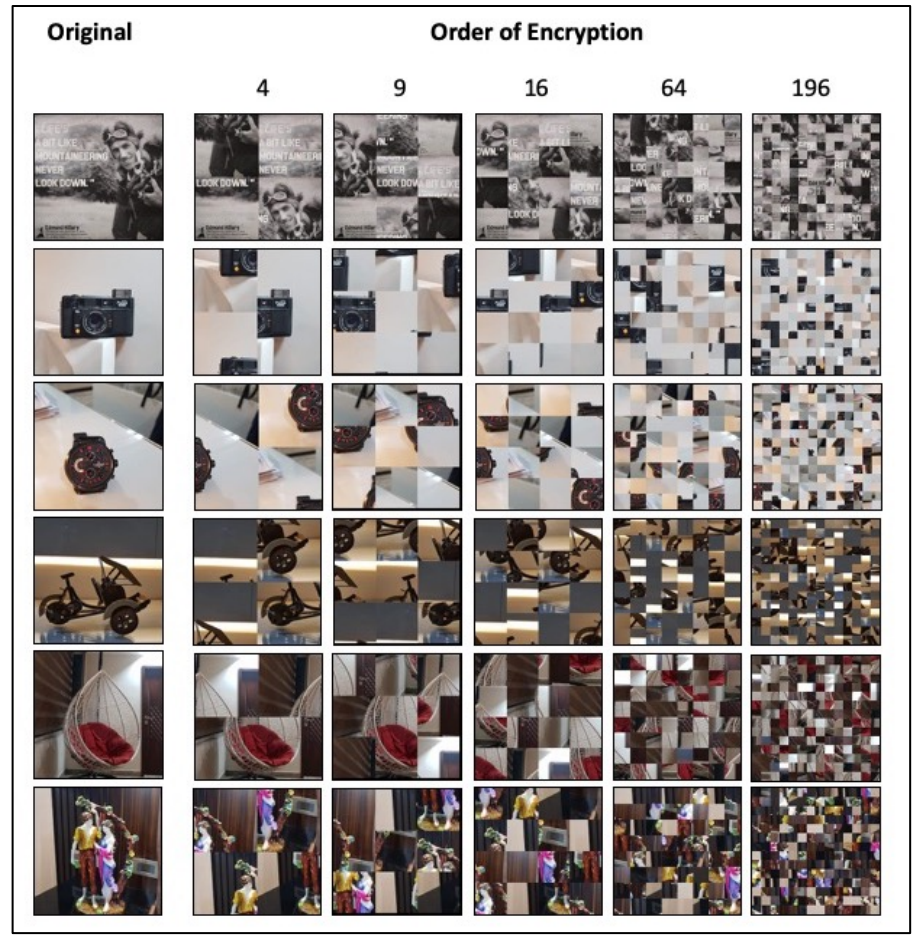

**Figure 4.** Encryption order (increasing left to right) of secret image comparison chart

## 3.5. Integrating Neural Networks with Encryption

### 3.5.1 Experimental Configuration

ML-Engine from Google Cloud Platform is used to train the model. JupyterLab on AI Platform's Notebook Instances which is currently in the β phase is used. In particular, we used Tensorflow 2.0 Notebook Instance that has pre-installed Tensorflow with Keras support. It also had all other essential artificial intelligence and machine learning libraries compiled in it. Model was developed using libraries including Tensorflow, Keras, Matplotlib, Numpy, Scipy, etc.

**Table 1.** Software and Hardware Specifications

| COMPONENT | SPECIFICATIONS |
|-----------|----------------|
| GPU | 1 NVIDIA Tesla K80 having 2496 CUDA(version 10.0) Cores, 12GB(11.46 Usable) DDR5 VRAM |
| CPU | 4 Intel Xeon CPU @ 2.20GHz |
| RAM | 15GB |
| Boot disk | 100 GB storage space available |
| Language | Python 3.5.3 |
| OS | Debian GNU/Linux 9.11 |

### 3.5.2. Dataset Preparation and Encryption

The Flickr30k dataset has been used to train the model. The images were of irregular size, therefore were scaled at 256x256 as per our training model. A total of 31,783 images were used which had an RGB scale. Datasets can be found at [41]. Furthermore, each image of the dataset is transformed into blocks that are scrambled thereafter.

### 3.5.3. Train Neural Net

To train the convolutional neural network, the β values being 0.25, 0.75 and 1 as represented in Table 5 have been used. A batch size of 32 images with 1000 epochs were used to train the model. Neural network took with the above mentioned hardware took about 8 hours to train for a particular beta value. We used 3 nodes (3x3, 4x4 and 5x5) in every layer. 2 nodes (conv_prep0 and conv_prep1) were used in the preparation network. Whereas 5 layers (conv_hide0, conv_hide1, conv_hide2, conv_hide3, conv_hide4) were used in hiding network as represented in Table 3. For reveal network, 5 layers (conv_rev0, conv_rev1, conv_rev2, conv_rev3, conv_rev4) as well which were integrated with Adam optimizer was used as shown in Table 4. The following layers are summarized briefly in Table 2. Retrieving the encrypted image can be revealed using the decryption methodology explained under the heading Image Encryption.

**Table 2.** Summary of the proposed neural network

| Model: "model_1" | | | |
|------------------|------------------|---------|-----------------------|
| Layer(type) | Output Shape | Param # | Connected to |
| input_1 (InputLayer) | (None, 256, 256, 3) | 0 | |
| input_2 (InputLayer) | (None, 256, 256, 3) | 0 | |
| Encoder (Model) | (None, 256, 256, 3) | 293273 | input_1[0][0] input_2[0][0] |

| Decoder (Model) | (None, 256, 256, 3) | 195388 | Encoder[1][0] |
|---|---|---|---|
| concatenate_14 (Concatenate) | (None, 256, 256, 6) | 0 | Decoder[1][0] Encoder[1][0] |
| Total params: 488,661 Trainable params: 293,273 Non-trainable params: 195,388 | | | |

**Table 3.** Detailed representation of the Encoder model

| Model: "Encoder" | | |
|---|---|---|
| Layer(type) | Output Shape | Connected to |
| input_3 | (None, 256, 256, 3) | |
| conv_prep0_(3x3 4x4 and 5x5) | (None, 256, 256, (50, 10, 5) ) | input_3[0][0] |
| concatenate_1 | (None, 256, 256, 65) | conv_prep0_(3x3, 4x4, 5x5)[0][0] |
| conv_prep1_(3x3, 4x4 and 5x5) | (None, 256, 256, (50, 10, 5) ) | concatenate_1[0][0] |
| input_4 | (None, 256, 256, 3) | |
| concatenate_2 | (None, 256, 256, 65) | conv_prep1_(3x3, 4x4 and 5x5)[0][0] |
| concatenate_3 | (None, 256, 256, 68) | input_4[0][0], concatenate_2[0][0] |
| conv_hid0_(3x3, 4x4, 5x5) | (None, 256, 256, (50, 10, 5)) | concatenate_3[0][0] |
| concatenate_4 | (None, 256, 256, 65) | conv_hid0_(3x3, 4x4, 5x5)[0][0] |
| conv_hid1_(3x3, 4x4, 5x5) | (None, 256, 256, (50, 10, 5)) | concatenate_4[0][0] |
| concatenate_5 | (None, 256, 256, 65) | conv_hid1_(3x3, 4x4, 5x5)[0][0] |
| conv_hid2_(3x3, 4x4, 5x5) | (None, 256, 256, (50, 10, 5)) | concatenate_5[0][0] |
| concatenate_6 | (None, 256, 256, 65) | conv_hid2_(3x3, 4x4, 5x5)[0][0] |
| conv_hid3_(3x3, 4x4, 5x5) | (None, 256, 256, (50, 10, 5)) | concatenate_6[0][0] |
| concatenate_7 | (None, 256, 256, 65) | conv_hid3_(3x3, 4x4, 5x5)[0][0] |

| conv_hid4_(3x3, 4x4, 5x5) | (None, 256, 256, (50, 10, 5)) | concatenate_7[0][0] |
|---|---|---|
| concatenate_8 | (None, 256, 256, 65) | conv_hid4_(3x3, 4x4, 5x5)[0][0] |
| output_C (Conv2D) | (None, 256, 256, 3) | concatenate_8[0][0] |
| Total params: 293,273<br>Trainable params: 293,273<br>Non-trainable params: 0 | | |

**Table 4.** Detailed representation of the Decoder model

| Model: "Decoder" | | |
|---|---|---|
| Layer(type) | Output Shape | Connected to |
| input_5 (Input Layer) | (None, 256, 256, 3) | |
| output_C_noise (GaussianNoise) | (None, 256, 256, 3 ) | input_5[0][0] |
| conv_rev0_(3x3, 4x4 and 5x5) | (None, 256, 256, (50,10,5) | output_C_noise[0][0] |
| concatentate_9 | (None, 256, 256, 65) | conv_rev0_(3x3, 4x4 and 5x5)[0][0] |
| conv_rev1_(3x3, 4x4 and 5x5) | (None, 256, 256, (50,10,5) | concatenate_9[0][0] |
| concatentate_10 | (None, 256, 256, 65) | conv_rev1_(3x3, 4x4 and 5x5)[0][0] |
| conv_rev2_(3x3, 4x4 and 5x5) | (None, 256, 256, (50,10,5) | concatenate_10[0][0] |
| concatentate_11 | (None, 256, 256, 65) | conv_rev2_(3x3, 4x4 and 5x5)[0][0] |
| conv_rev3_(3x3, 4x4 and 5x5) | (None, 256, 256, (50,10,5) | concatenate_11[0][0] |
| concatentate_12 | (None, 256, 256, 65) | conv_rev3_(3x3, 4x4 and 5x5)[0][0] |
| conv_rev4_(3x3, 4x4 and 5x5) | (None, 256, 256, (50,10,5) | concatenate_12[0][0] |
| concatentate_13 | (None, 256, 256, 65) | conv_rev4_(3x3, 4x4 and 5x5)[0][0] |
| output_S (Conv2D) | (None, 256, 256, 3) | concatenate_13[0][0] |
| Total params: 390,776<br>Trainable params: 195,388<br>Non-trainable params: 195,388 | | |

## 4. Results

**Table 5.** MSE per pixel each for cover and secret.

| β | COVER | SECRET |
|---|---|---|
| **1** | 13.8137 | 11.2884 |
| **0.75** | 12.1389 | 13.6526 |
| **0.25** | 8.81926 | 17.5474 |

Table 5 represents the mean of Sum of Squared Errors per pixel each for cover and secret images. According to which β = 1 shows the minimum error in the secret image which is beneficial since data lost in-network is minimum in this case and hence chosen as the optimum value. Though it has maximum pixel error in the cover image, the difference from other β values is not as significant as seen in pixel errors of the secret image.

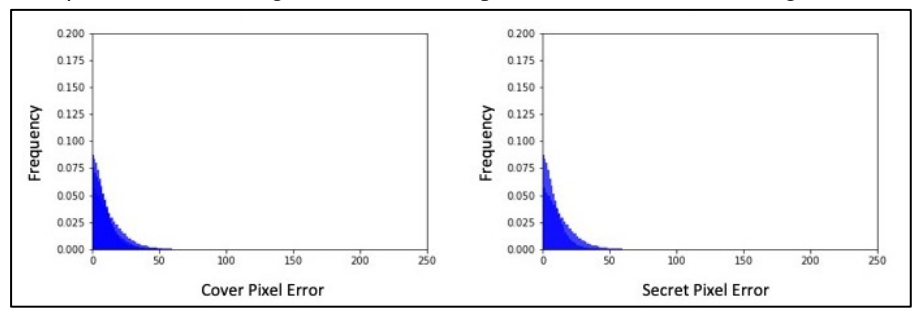

**Figure 5.** For β = 1, Left: Cover pixel error histogram, Right: Secret Pixel Error

The above graphs shown in Figure 5 represent the distribution of error in each pixel of both cover and secret image after encoding and decoding process for the most appropriate value of β i.e. 1.

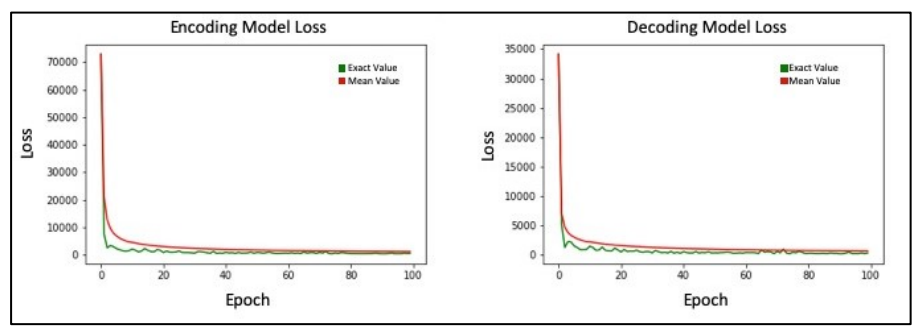

**Figure 6.** Left: Encryption Model Loss, Right: Decryption Model Loss

Figure 6 depicts the mean values for β = 1. The green and red on the line plot represent exact values and mean values respectively. Encoding Loss is the total loss of model that is the sum of squares of |C-C'| + β*|S-S'|. Reveal loss is the sum of squares of β*|S-S'|. This methodology is majorly dependent upon auto-encoders. Nonetheless, instead of simply encoding an image, two images are convoluted making the cover and container analogous to each other. Consequently, β defines the reconstruction error.

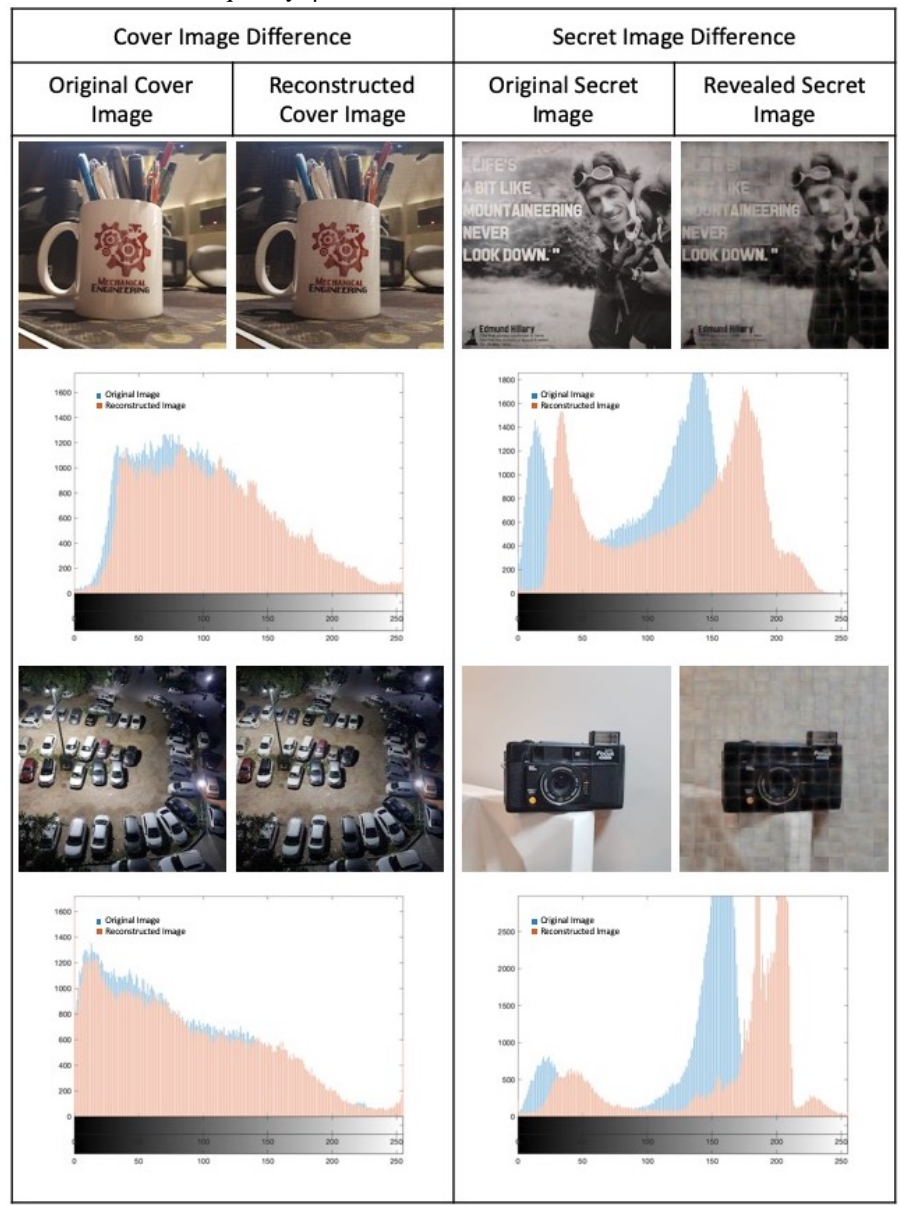

**Figure 7.** Representation of pixel densities in cover and secret image using histograms.

The histograms in Figure 7 represent the difference in pixel density in the original cover (left side) and secret (right side) images. Orange color represents the original cover and secret whereas blue color represents the reconstructed cover and revealed secret respectively. Due to adding an extra layer for cryptography, there is a tradeoff between secrecy and reveal loss.

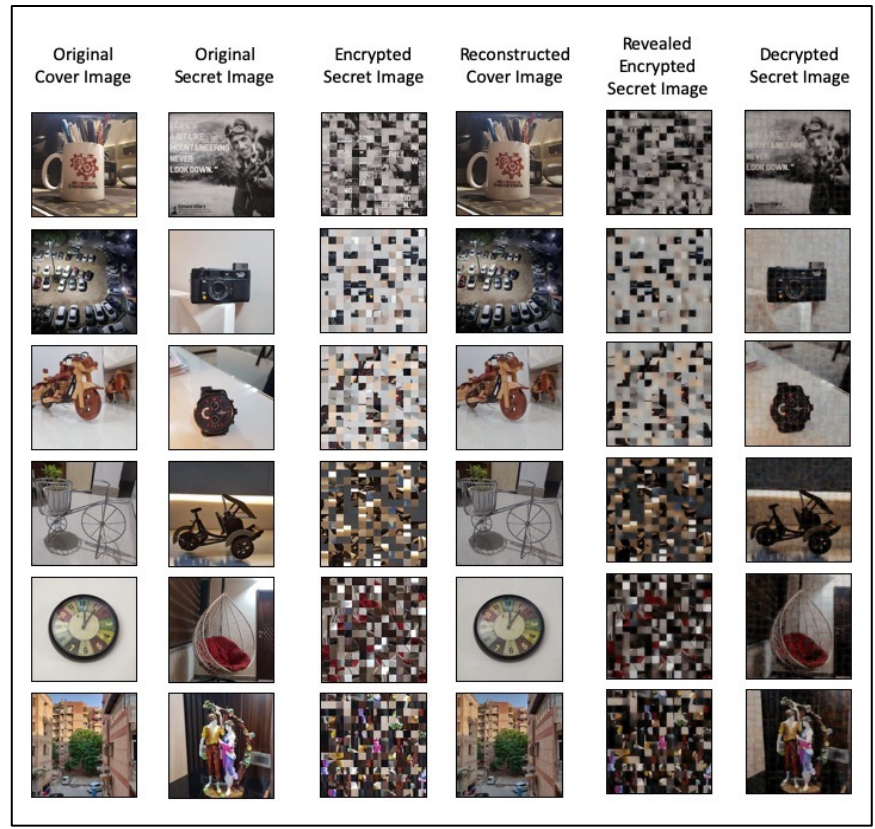

**Figure 8.** Results of integrating neural networks with an encryption layer. Left pair: original cover and secret image, Center pair: encrypted secret and reconstructed cover image, Right pair: revealed encrypted and decrypted the secret image.

As shown in Figure 8, the first column represents the original cover image used to hide the encrypted secret image shown in the third column. The fourth column represents the reconstructed cover image which has an embedded encrypted secret image and we finally get the decrypted secret image in the sixth column. The output represents that the secret image is successfully retrieved. Although, it contains some error the secret image is easily understandable. Figure 8 showcases only 6 tuples, but we were successfully able to hide and retrieve over 1000 images.

## 5. What If the Original Cover Image Is Available?

As explained by Baluja [11], the original cover image i.e. without the secret image embedded is almost impossible for someone to obtain. But what if it was obtained? After examining, without the decoding network secret could be partially be revealed. The secret image can be revealed by taking the difference between the original and reconstructed image i.e. (|S-S'|).

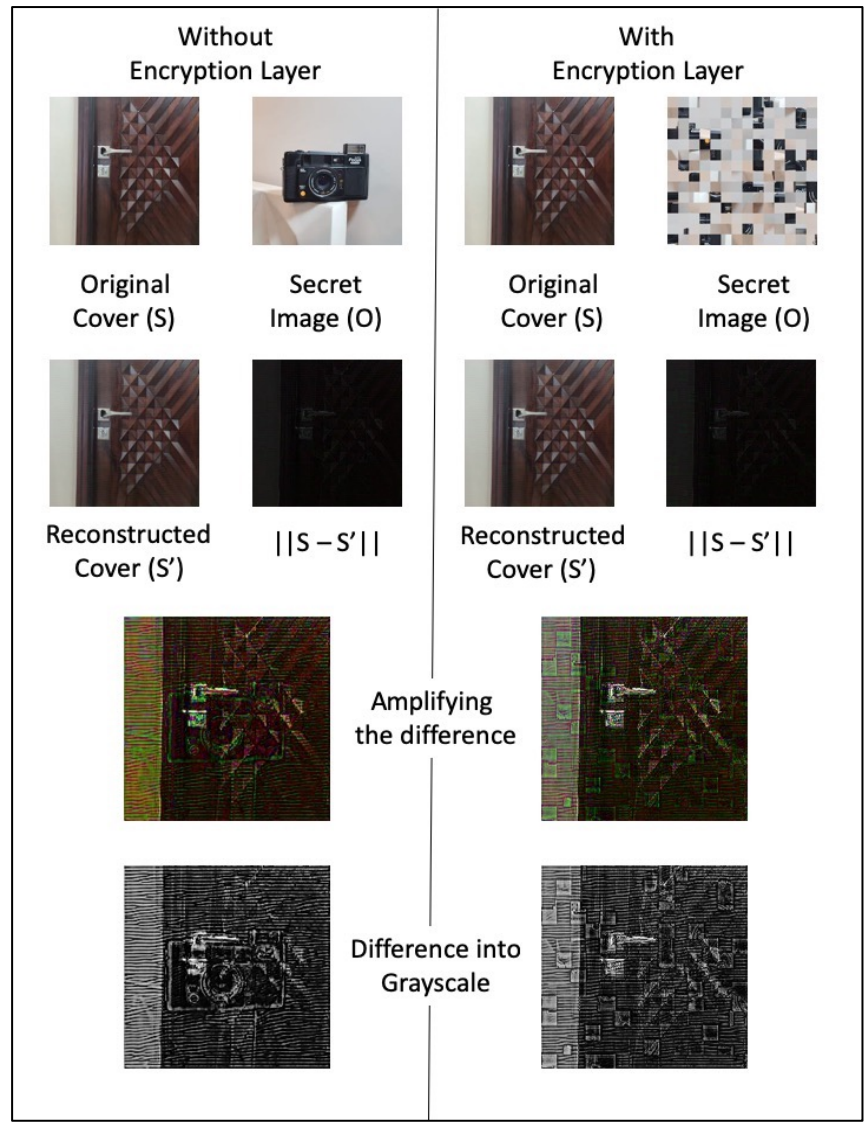

**Figure 9.** Effect of encryption layer.

As Figure 9 shows, without the first layer of encryption, data can be revealed by taking the difference between the original cover image and the reconstructed image. After obtaining the difference the secret is not visible but after enhancing the image and converting it into grayscale, the great resemblance between secret image and the residual image is found.

If we use the first layer of encryption, the secret is impossible to obtain as after enhancing and converting the residual image into grayscale we see only boxes and the residual image is not comprehensible which is depicted in Figure 9. Hence, it makes the system robust and almost impossible for someone to extract the features of the secret image.

## 6. Conclusion

This paper propounds a new method for image steganography which is much more secure than the previous implementations. Although, it uses some common steganography techniques, integrating it with cryptography and neural networks make it arduous to break. The encryption layer added provides an additional layer of security with deep neural networks. As shown in Figure s8 we successfully embedded and revealed the secret image from the container. Further, the main reason to add an additional encryption layer was that if the original cover was made public, the secret will still remain secure. In earlier implementations, as shown in Figure 9 if the additional layer is not added, one can partially decrypt the sensitive information.

## 7. Future Work

In this section, we discuss possibilities for future work. Primarily, the method used for encryption can be replaced with some sophisticated encryption algorithms. Although the method for encrypting the secret image is secure, using cryptographic methods such as Advanced Encryption Standard (AES) or Data Encryption Standard (DES) is beyond the scope of this paper. The reason behind not using those algorithms is the problem faced when recovering the encrypted image due to lossy neural networks. In the future, if lossless neural networks are achieved, we would be able to implement them with modern cryptography techniques. The scope of building cryptographic algorithms that could work with neural networks is also a great field to explore in the future.

Moreover, integrating neural networks with image steganography creates a variety of possibilities in this domain. As stated by Gopalan (2002) [42] LSB steganography can be implemented with audio files as well. Therefore, we can similarly integrate neural nets with audio files. Not only LSB steganography there are numerous other steganography techniques with which neural networks can be integrated in the future.

## Conflicts of Interest

There is no conflict of interest.